\documentclass[a4paper,12pt]{article}
\usepackage{epsfig}
\usepackage{citesort}
\usepackage{psfrag}
\date{\today}
 \normalsize

\def\be{\begin{equation}}
\def\ee{\end{equation}}
\def\bea{\begin{eqnarray}}
\def\eea{\end{eqnarray}}

\def\lsim{\raise0.3ex\hbox{$\;<$\kern-0.75em\raise-1.1ex\hbox{$\sim\;$}}}
\def\gsim{\raise0.3ex\hbox{$\;>$\kern-0.75em\raise-1.1ex\hbox{$\sim\;$}}}

\def\ie{{\it i.e.}}

\def\no{\nonumber\\}

\begin{document}
\renewcommand{\thefootnote}{\fnsymbol{footnote}}
\rightline{\today} \vspace{.3cm} {\large
\begin{center}
{\bf Supersymmetric contributions to $B\to K \pi$ in the view of
recent experimental results}
\end{center}}
\vspace{.3cm}
\begin{center}
Shaaban Khalil\\
\vspace{.3cm}%
\emph{Center for Theoretical Physics, British University in Egypt,
Shorouk city, \\
Cairo, 11837, Egypt.}\\
\emph{Ain Shams University, Faculty of Science, Cairo, 11566,
Egypt.}
\end{center}
\vspace{.3cm} \hrule \vskip 0.3cm
\begin{center}
\small{\bf Abstract}\\[3mm]

\begin{minipage}[h]{15.2cm}
Supersymmetric contributions to the branching ratios and CP
asymmetries of $B\to K \pi$ decays are analyzed in the view of
recent experimental measurements. We show that supersymmetry can
still provide a natural solution to the apparent discrepancy
between theses results and the standard model expectations. We
emphasize that chargino contributions may enhance the electroweak
penguin effects that can resolve to the $B\to K\pi$ puzzle. We
also point out that a non-universal $A$-terms is an essential
requirement for this solution.

\end{minipage}
\end{center}
\vspace{.3cm} \hrule \vskip 0.5cm
%
\section{\large{\bf Introduction}}
Recently, BaBar \cite{Aubert:2006ap} and Belle \cite{belle}
collaborations have reported new experimental results for the
branching ratios (BRs) and CP asymmetries of $B \to K \pi$ decays.
As in the previous measurements \cite{Aubert:2004qm}, the current
results point to a lack of compatibility with the Standard Model
(SM) expectations, which in known as the $B\to K \pi$ puzzle. The
new average values of the experimental measurements of the BRs and
CP asymmetry of the four decay channels \cite{Gronau:2006xu} are
given in Table 1.

\begin{table}[h]
\begin{center}
\begin{tabular}{|c|c|c|}
  \hline
  \hline
  Decay channel & $BR \times 10^{-6}$ & $A_{\scriptscriptstyle CP}$ \\
  \hline
  \hline
  $K^+ \pi^-$ & $19.83\pm 0.63$ & $-0.099\pm 0.016$ \\
  $K^+ \pi^0$ & $12.83\pm 0.59$ & $0.050\pm 0.025$ \\
  $K^0 \pi^+$ & $23.4 \pm 1.06$ & $0.007\pm 0.025$ \\
  $K^0 \pi^0$ & $9.89\pm 0.63$ & $-0.12\pm 0.11$ \\
  \hline
\end{tabular}
\vskip 0.25cm \caption{The new average results for the BRs and CP
asymmetries of $B \to K \pi$ decays.}
\end{center}
\end{table}

It is important to notice that the updated value of the direct CP
asymmetry in $B^0 \to K^+ \pi^-$, $A_{\scriptscriptstyle CP}(
K^+\pi^-) =-0.099\pm 0.016$, corresponds to a $4.3~ \sigma$
deviation from zero. Also the difference between the CP
asymmetries $A_{\scriptscriptstyle CP}(K^+\pi^-)$ and
$A_{\scriptscriptstyle CP}(K^+\pi^0)$ is about $3.2~\sigma$, which
is quite difficult to be accommodated within the SM, as emphasized
in Ref.\cite{Khalil:2005qg}. These results might indicate to a
large color-suppressed amplitude or enhanced electroweak penguin
as it happens in the supersymmetric (SUSY) extension of the SM
\cite{Khalil:2005qg,Mishima:2004um,Khalil:2004yb}.

From the latest measurements for BRs, one finds that the ratios
$R_c, R_n$ and $R$ of $B \to K \pi$ decays are given by
\begin{eqnarray} %
R_c &=&2\left[\frac{BR(B^+\to K^+\pi^0)+BR(B^-\to K^- \pi^0)}
{BR(B^+\to K^0 \pi^+)+ BR(B^- \to \bar{K}^0 \pi^-)} \right] =  1.096\pm 0.071,\label{Rcresult}\\
R_n & = & \frac{1}{2} \left[\frac{BR(B^0 \to K^+ \pi^-) +
BR(\bar{B}^0 \to K^- \pi^+)}{BR(B^0 \to K^0 \pi^0) + BR(\bar{B}^0
\to \bar{K}^0 \pi^0)} \right]=1.003\pm 0.071, \label{Rnresult}\\
R&=&\frac{\tau_B^+} {\tau_{B^0}}~\left[\frac{BR(B^0\to K^+\pi^-) +
BR(\bar{B}^0 \to K^- \pi^+)} {BR(B^+\to K^0 \pi^+)+ BR(B^- \to
\bar{K}^0 \pi^-)} \right]=0.923\pm 0.051. \label{Rresult}
\end{eqnarray}%
It is remarkable that $R_n$ has changed significantly from the
previous result, where $R_n$ was given by $R_n=0.79\pm 0.08$. It
is now very close to one, which makes it more consistent with the
SM and SUSY expectations. As discussed in
Ref.\cite{Khalil:2005qg}, it was rather difficult to account for
the situation $R_n < 1$ and $R_c \gsim 1$ in both of the SM and
SUSY models. In the SM, the amplitudes of $B\to K\pi$ imply that
$R_n = R_c \simeq 1$. While in SUSY models, it is possible to have
a deviation between $R_n$ and $R_c$ and to get $R_n$ less than
one. However, it has been realized
\cite{Khalil:2005qg,Khalil:2004yb} that it is quite unnatural to
obtain $R_n \simeq 0.79$ with $R_c\simeq 1.1$, although this may
occur in a very small region of the parameter space, as shown in
Fig. 2 in Ref.\cite{Khalil:2005qg}.

In this letter we update our previous analysis for the
supersymmetric contributions to the $B\to K \pi$ process. We show
that the possible supersymmetric solution to the $B\to K \pi$
puzzle is still consistent with the new experimental results. In
fact with the new measurements for the branching ratios, it is now
even easier for SUSY to account for the CP asymmetries of $B\to K
\pi$ decays in a wider region of the parameter space. We point out
that chargino contributions can enhance the elctroweak penguin
effects and account for the new experimental results of the BRs
and CP asymmetries. We indicate that the left-right (LR) mixing
between the second and third generation of up-squarks can provide
the source of the required flavor violation in this process.

The article is organized as follows. In section 2 we analyze the
supersymmetric contribution to the BRs of the $B\to K \pi$ decays.
We show that with the new experimental results, the suprsymmetric
solution to the difference $R_n-R_c$ can take place for more
points in the parameter space than before. In section 3 we study
the superysmmetric contribution to the CP asymmetries of $B\to K
\pi$ decays in view of the recent experimental results. In section
4 we briefly discuss the possibility of having a large mixing in
supersymmeric models. Finally we give our conclusions in section
5.

%
\section{\large{\bf SUSY contribution to $B\to K\pi$ branching ratios }}
The $B\to K\pi$ decays are driven by the $b \to s $ transition. In
supersymmetric theories, the effective Hamiltonian of this
transition is given by
\begin{eqnarray}
H^{\Delta B=1}_{\rm eff}&\!=\!&\frac{G_F}{\sqrt{2}} \sum_{p=u,c}
\lambda_p \Big(C_1 Q_1^p + C_2 Q_2^p + \sum_{i=3}^{10}C_i Q_i +
C_{7\gamma} Q_{7\gamma} + C_{8g} Q_{8g}\Big)  +h.c.,~~
\label{Heff}
\end{eqnarray}
where $\lambda_p= V_{pb} V^*_{ps}$, $Q_i$ are the relevant local
operators and $C_i$ are the Wilson coefficients  which can be
found in Ref.\cite{Khalil:2005qg}. The decay amplitudes of $B\to K
\pi$ can be parameterized as follows: %
\bea %
A_{B^+\to K^0 \pi^+}&=&\lambda_c A_{\scriptscriptstyle \pi K}
P\left[e^{-i\theta_P}+r_A e^{i\delta_A}e^{i\gamma}\right]
\label{susypar1} \\
\sqrt{2}A_{B^+\to K^+ \pi^0 }&=&\lambda_c A_{\scriptscriptstyle
\pi K} P\big[e^{-i\theta_P}+ \left(r_A e^{i\delta_A} +
r_{\scriptscriptstyle C} e^{i\delta_{\scriptscriptstyle C}}\right)
e^{i\gamma}+ r_{\scriptscriptstyle EW}
e^{-i\theta_{\scriptscriptstyle EW}}
e^{i\delta_{\scriptscriptstyle EW}}
\big] \label{susypar2}\\
A_{B^0\to K^+ \pi^- }&=& \lambda_cA_{\scriptscriptstyle \pi
K}P\left[e^{-i\theta_{\scriptscriptstyle
P}}+\left(r_{\scriptscriptstyle A} e^{i\delta_{\scriptscriptstyle
A}}+ r_{\scriptscriptstyle T} e^{i\delta_{\scriptscriptstyle
T}}\right)e^{i\gamma} + r_{\scriptscriptstyle
EW}^{\scriptscriptstyle C} e^{-i\theta_{\scriptscriptstyle EW}}
e^{i\delta_{\scriptscriptstyle EW}^{\scriptscriptstyle C}}\right], \label{susypar3}\\
-\sqrt{2}A_{B^0\to K^0 \pi^0}&\!=\!&\lambda_c
A_{\scriptscriptstyle \pi K} P \Big[e^{-
i\theta_{\scriptscriptstyle P}}\!+\! \left( r_{\scriptscriptstyle
A} e^{i\delta_{\scriptscriptstyle A}} + r_{\scriptscriptstyle T}
e^{i\delta_{\scriptscriptstyle T}} - r_{\scriptscriptstyle C}
e^{i\theta_{\scriptscriptstyle C}}e^{i\delta_{\scriptscriptstyle
C}}\right) e^{i\gamma} + r_{\scriptscriptstyle
EW}^{\scriptscriptstyle C} e^{- i\theta_{\scriptscriptstyle
EW}^{\scriptscriptstyle C}}e^{i\delta_{\scriptscriptstyle
EW}^{\scriptscriptstyle C}}
\nonumber\\
&-& r_{\scriptscriptstyle EW}
e^{- i\theta_{\scriptscriptstyle EW}}e^{i\delta_{\scriptscriptstyle EW}}\Big],%
\label{susypar4}%
\eea%
where $\delta_{\scriptscriptstyle A}, \delta_{\scriptscriptstyle
C}, \delta_{\scriptscriptstyle T}, \delta_{\scriptscriptstyle EW},
\delta_{\scriptscriptstyle EW}^{\scriptscriptstyle C}$ and
$\theta_{\scriptscriptstyle P}, \theta_{\scriptscriptstyle EW},
\theta_{\scriptscriptstyle EW}^{\scriptscriptstyle C}$ are the CP
conserving (strong) and the CP violating phase, respectively. Here
${\small T,C,A,P, EW, EW^C}$ represent a tree, a color suppressed
tree, an annihilation, QCD penguin, electroweak penguin, and
suppressed electroweak penguin diagrams, respectively. The
parameters
$P, r_{\scriptscriptstyle EW}, r_{\scriptscriptstyle EW}^{\scriptscriptstyle C}$ are defined as %
\bea %
P e^{i \theta_{\scriptscriptstyle P}}
e^{i\delta_{\scriptscriptstyle
P}}&=&\alpha_4^c-\frac{1}{2}\alpha^c_{\scriptscriptstyle
4,EW}+\beta_3^c+\beta^c_{\scriptscriptstyle 3,EW}~
,\no %
r_{\scriptscriptstyle EW} e^{i \theta_{\scriptscriptstyle EW}}
e^{i\delta_{\scriptscriptstyle
EW}}&=&\left[\frac{3}{2}(R_{\scriptscriptstyle
K\pi}\alpha^c_{\scriptscriptstyle 3, EW}+
\alpha^c_{\scriptscriptstyle 4,EW})\right]/P~,\no%
r^C_{\scriptscriptstyle EW} e^{i \theta_{\scriptscriptstyle
EW}^C}e^{i\delta_{\scriptscriptstyle EW}^C}&=&
\left[\frac{3}{2}(\alpha^c_{\scriptscriptstyle 4,EW}-\beta^c_{\scriptscriptstyle 3,EW})\right]/P~.%
\eea %
Detailed definitions for these parameters in terms of the relevant
Wilson coefficients of the QCD and electroweak penguins can be
found in Ref.\cite{Khalil:2005qg}. In this case, one can expand
$R_c$ and $R_n$ in terms of $r_{\scriptscriptstyle T}$,
$r_{\scriptscriptstyle EW}$ and $r^{\scriptscriptstyle
C}_{\scriptscriptstyle EW}$ as follows
\cite{Khalil:2005qg}: %
\bea %
R_c &\simeq& 1 + r_{\scriptscriptstyle T}^2 -
2r_{\scriptscriptstyle T}\cos (\gamma + \theta_{\scriptscriptstyle
P}) + 2r_{\scriptscriptstyle EW}\cos (\theta_{\scriptscriptstyle
P} - \theta_{\scriptscriptstyle EW}) - 2r_{\scriptscriptstyle T}
r_{\scriptscriptstyle EW}\cos (\gamma
 + \theta_{\scriptscriptstyle EW}),~~~~~\\
R_c-R_n &\simeq& 2r_{\scriptscriptstyle T} r_{\scriptscriptstyle
EW}\cos (\gamma +2\theta_{\scriptscriptstyle
P}-\theta_{\scriptscriptstyle EW}) -2r_{\scriptscriptstyle T}
r_{\scriptscriptstyle EW}^{\scriptscriptstyle C}\cos (\gamma
+2\theta_{\scriptscriptstyle EW}-\theta_{\scriptscriptstyle
EW}^{\scriptscriptstyle C}),
\label{eq:RcRn}%
\eea %
where $r_{\scriptscriptstyle C}\simeq r_{\scriptscriptstyle T}$
has been assumed \cite{Khalil:2005qg}. It interesting to note that
within the SM, where $\theta_{\scriptscriptstyle  P}=
\theta_{\scriptscriptstyle EW}=\theta_{\scriptscriptstyle
EW}^{\scriptscriptstyle C}=0$ and $r_{\scriptscriptstyle
EW}^{\scriptscriptstyle C} \simeq 0.01 << r_{\scriptscriptstyle
EW}\simeq
0.1$, one obtains%
\be %
(R_c -R_n) \Big\vert_{\scriptscriptstyle SM} \simeq {\cal O}(0.01)~, %
\ee %
which is not consistent with the experimental results.

For gluino mass $m_{\tilde{g}}=500$ GeV, average squark mass
$m_{\tilde{q}}=500$ GeV, light stop mass $m_{\tilde{t}_R}=150$
GeV, $M_2=200$ GeV, $\mu=400$ GeV and $\tan \beta=10$, the SUSY
contributions to the $r_{\scriptscriptstyle EW}$ and
$r_{\scriptscriptstyle EW}^{\scriptscriptstyle
C}$ are given by \cite{Khalil:2005qg} %
\be %
r_{\scriptscriptstyle EW}^{\scriptscriptstyle SUSY} \simeq
r_{\scriptscriptstyle EW}^{\scriptscriptstyle SM} \left[ 1+ 0.053
\tan\beta (\delta^u_{LL})_{32} -2.78 (\delta^d_{LR})_{23} + 1.11
(\delta^u_{LR})_{32} \right]. %
\label{eq-rew}%
\ee%
and
\be %
(r_{\scriptscriptstyle EW}^{\scriptscriptstyle
C})^{\scriptscriptstyle SUSY} \simeq (r_{\scriptscriptstyle
EW}^{\scriptscriptstyle C})^{\scriptscriptstyle SM} \left[ 1+
0.134 \tan\beta (\delta^u_{LL})_{32} +26.4 (\delta^d_{LR})_{23} +
1.62 (\delta^u_{LR})_{32} \right]. %
\label{eq-rewc}%
\ee%
It is worth mentioning that the $r_{\scriptscriptstyle EW}$ and
$r_{\scriptscriptstyle EW}^{\scriptscriptstyle C}$ dependence on
the down mass insertion $(\delta^d_{LR})_{23}$ is due to the
gluino contribution, mainly via the chromomagnetic operator
$Q_{8g}$. While the up mass insertions $(\delta^u_{LL})_{32}$ and
$(\delta^u_{LR})_{32}$ are due to the chargino contributions
(penguin diagrams with chargino in the loop), in particular
through $Q_{7\gamma}$, $Q_{8g}$ and $Z$-penguin, for more details
see Ref.\cite{Khalil:2005qg}.

The mass insertion $(\delta^d_{LR})_{23}$ is constrained by the
branching ratio of $b \to s \gamma$, such that $\vert
(\delta^d_{LR})_{23}\vert \lsim 10^{-2}$. Also the mass insertion
$(\delta^u_{LL})_{32}$ can be restricted by $b\to s \gamma$,
however this constraints depend on the value of $\tan \beta$ and
the phase of this mass insertion, as discussed in details in
Ref.\cite{Khalil:2005qg}. Finally, the $(\delta^u_{LR})_{32}$ mass
insertion is essentially unconstrained and can be of order one.
Therefore, as can be seen from Eq.(\ref{eq-rewc}), $
r_{\scriptscriptstyle EW}^{\scriptscriptstyle C}$ can not be
enhanced much in SUSY models and its typical value is of order
${\cal O}(10^{-2})$ as in the SM, especially in the scenario of
dominant mass insertion $(\delta^u_{LR})_{32}$, which we will
adopt in our analysis.

In this case, it is quite safe to neglect the effect of
$r_{\scriptscriptstyle EW}^{\scriptscriptstyle C}$ respect to
$r_{\scriptscriptstyle EW}$. Therefore, the difference $R_c -R_n$
can be of order $0.1$ if $r_{\scriptscriptstyle T}
r_{\scriptscriptstyle EW}$ is of order $0.05$ and $\cos(\gamma + 2
\theta_{\scriptscriptstyle P} - \theta_{\scriptscriptstyle EW})
\simeq 1$. Since $r_{\scriptscriptstyle T}$ is dominated by the SM
values and it is given \cite{Khalil:2005qg} by
$r_{\scriptscriptstyle T}^{\scriptscriptstyle SM} \simeq 0.2$, a
value of order $0.25$ is required for $r_{\scriptscriptstyle EW}$
to have $R_c -R_n \simeq 0.1$. As can be seen from
Eq.(\ref{eq-rew}), such value of $r_{\scriptscriptstyle EW}$ can
be obtained with $(\delta^u_{LR})_{32} \sim {\cal O}(1)$. It is
important to mention that with the previous experimental result
for $R_c-R_n$ which was of order $0.2$, it was not possible to
saturate this difference with single mass insertion contribution
and simultaneous contributions from two mass insertions at least
are required \cite{Khalil:2004yb}.
%
\section{\large{\bf SUSY contribution to $B\to K\pi$ CP asymmetry }}
The direct CP asymmetry of $B^0 \to K^+ \pi^-$ decay is
defined as %
\be %
A_{CP}(K^+\pi^-) = \frac{\big\vert A(B^0 \to K^+
\pi^-)\big\vert^2- \big\vert A(\bar{B}^0 \to K^-
\pi^+)\big\vert^2}{ \big\vert A(B^0 \to K^+ \pi^-)\big\vert^2
+\big\vert A(\bar{B}^0 \to K^- \pi^+)\big\vert^2},%
\ee%
with similar expressions for the asymmetries
$A_{CP}(\bar{K}^0\pi^-)$, $A_{CP}(K^-\pi^0)$ and
$A_{CP}(\bar{K}^0\pi^0)$. As is known, a necessary condition to
generate a CP asymmetry is that the corresponding process should
have at least two interfering amplitudes with different weak and
strong phases. Using the above parametrization in
Eqs.(\ref{susypar1}-\ref{susypar4}), one finds %
\bea %
A_{CP}(K^+\pi^-) &\simeq & - 2 r_{\scriptscriptstyle T} \sin
\delta_{\scriptscriptstyle T} \sin(\theta_{\scriptscriptstyle P} +
\gamma),\label{CP1}\\
A_{CP}(K^0\pi^+) &\simeq & -  2 r_{\scriptscriptstyle A} \sin
\delta_{\scriptscriptstyle A} \sin(\theta_{\scriptscriptstyle P} + \gamma), \label{CP2}\\
A_{CP}(K^0\pi^0) &\simeq &  2 r_{\scriptscriptstyle EW}\sin
\delta_{\scriptscriptstyle EW}
\sin(\theta_{\scriptscriptstyle P}- \theta_{\scriptscriptstyle EW}),\label{CP3}\\
A_{CP}(K^+\pi^0) &\simeq &  - 2 r_{\scriptscriptstyle T} \sin
\delta_{\scriptscriptstyle T} \sin(\theta_{\scriptscriptstyle
P}+\gamma) - 2 r_{\scriptscriptstyle EW}\sin
\delta_{\scriptscriptstyle EW} \sin(\theta_{\scriptscriptstyle P}-
\theta_{\scriptscriptstyle EW}),\label{CP4}%
\eea %
where we have neglected $r_{\scriptscriptstyle
EW}^{\scriptscriptstyle C}$ respect to $r_{\scriptscriptstyle EW}$
and $r_{\scriptscriptstyle T}$. Also we ignored the higher order
terms. Here some comments, that can be concluded from the above
approximated expressions for the CP asymmetries, are in order:%
\begin{enumerate}%
\item%
Within the SM, where $\theta_{\scriptscriptstyle
P}=\theta_{\scriptscriptstyle EW}=0$, $r_{\scriptscriptstyle
T}\simeq 0.2$ and $r_{\scriptscriptstyle EW} \simeq 0.1$, one
finds that%
\be%
A_{CP}(K^+ \pi^0) = A_{CP}(K^+ \pi^-), %
\ee%
which is not supported by the recent data reported in Table 1.%
\item %
In the SM, the CP asymmetry $A_{CP}(K^0 \pi^0)$ is expected to be
close to zero. This may contradict the recent results indicate
that
$A_{CP}(K^0 \pi^0) \simeq -0.12\pm 0.11$.%
\item%
The CP asymmetry $A_{CP}(K^0 \pi^+)$ seems consistent with the SM
since $r_{\scriptscriptstyle A} \simeq {\cal O}(0.01)$.%
\item %
It is remarkable that the values of the SUSY CP violating phases
$\theta_{\scriptscriptstyle P}$ and $\theta_{\scriptscriptstyle
EW}$ would play important role in accommodating the experimental
measurements of these CP asymmetries.
\end{enumerate}

Now, let us discuss the SUSY contributions to the CP asymmetries
$A_{CP}(K\pi)$. As can be seen from Table 1, the CP asymmetry
$A_{CP}(K^0 \pi^0)$ measurement includes a large uncertainty.
Therefore, in our analysis for the supersymmetric contributions,
we will focus on $A_{CP}(K^+ \pi^-)$ and $A_{CP}(K^+ \pi^0)$.
Nonetheless, we will derive the corresponding $A_{CP}(K^0 \pi^0)$
in the region of the SUSY parameter space that leads to a
consistent results with experimental measurements. Concerning the
$A_{CP}(K^0 \pi^+)$, since $r_{\scriptscriptstyle A}$ receives
negligible SUSY contribution, it remains, as in the SM, of order
$0.01$. Thus, with a proper value of $\delta_{\scriptscriptstyle
A}$ one can easily get the measured small value $A_{CP}(K^+
\pi^0)$.
\begin{figure}[t]
\begin{center}
\epsfig{file=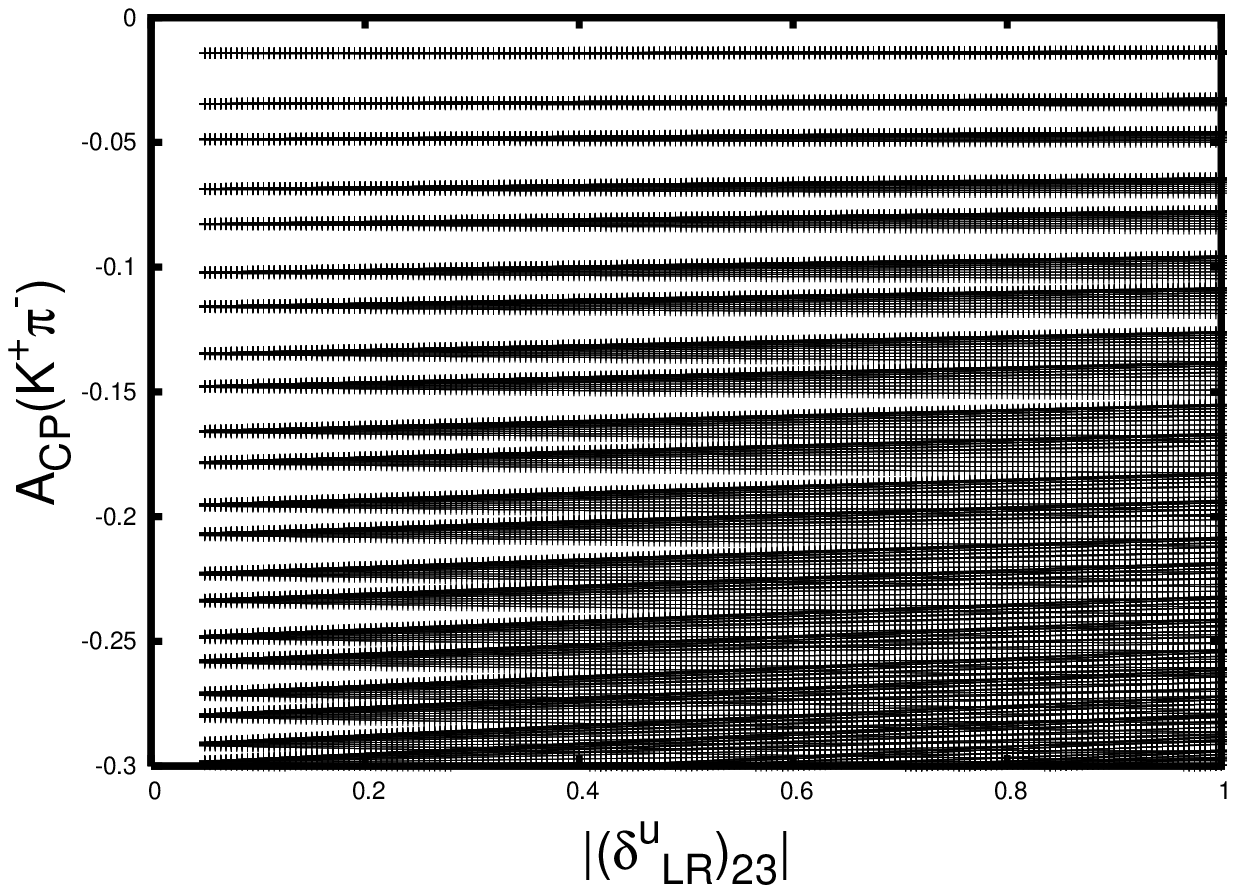,width=7cm,height=5.cm}\hspace{1cm}
\epsfig{file=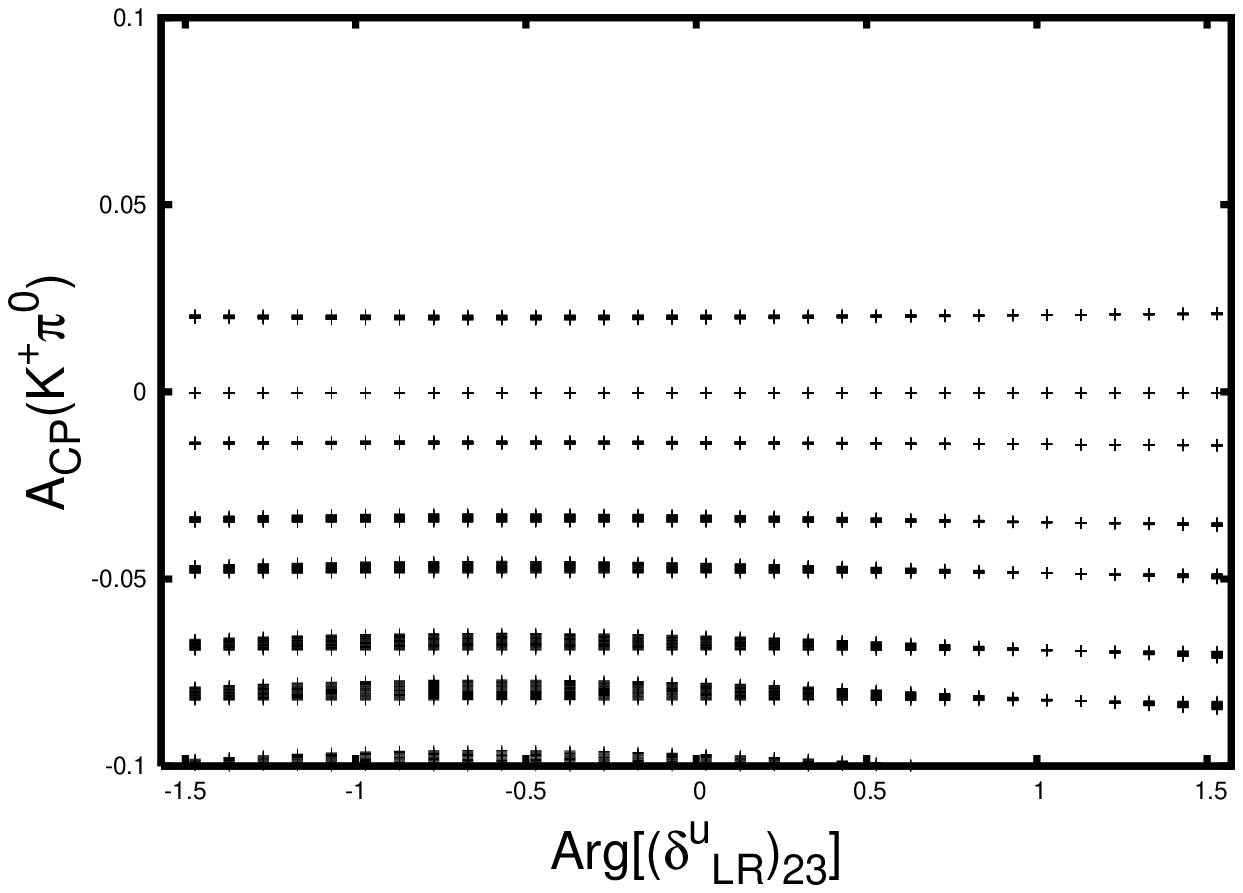,width=7cm,height=5.cm} %
\caption{(left) CP asymmetry $A_{CP}(K^+ \pi^-)$ as function of
$\vert(\delta^u_{LR})_{32}\vert$ for
$\rm{Arg}[(\delta^u_{LR})_{32}]\in[-\frac{\pi}{2},\frac{\pi}{2}]$.
(right) $A_{CP}(K^+ \pi^-)$ as function of
$\rm{Arg}[(\delta^u_{LR})_{32}]$ for
$\vert(\delta^u_{LR})_{32}\vert\in [0,1]$. The strong phase
$\delta_{\scriptscriptstyle  T}$ is varied in the range
$\left[-\pi,0\right]$.} \vspace{0.3cm}
\end{center}
\end{figure}
\begin{figure}[t]
\begin{center}
\epsfig{file=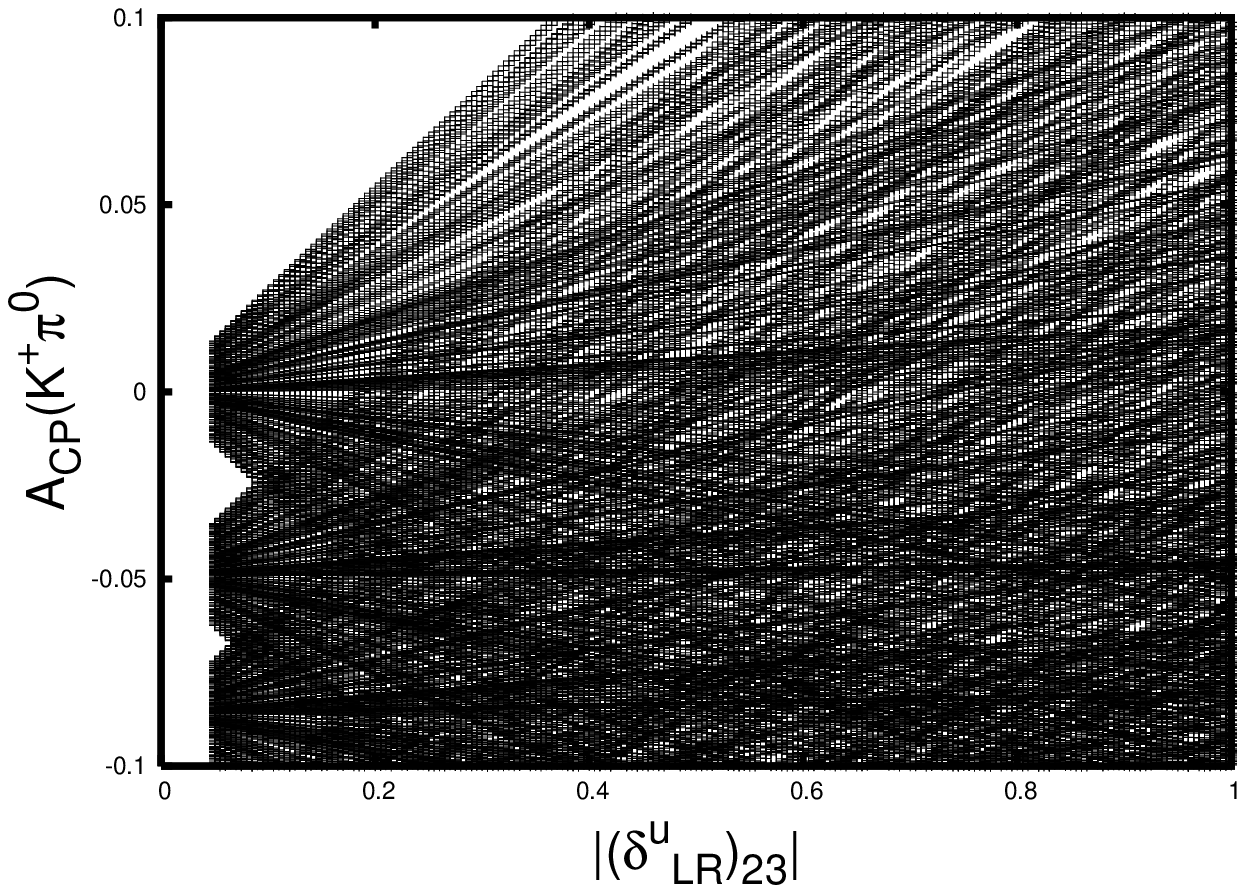,width=7cm,height=5.cm}\hspace{1cm}
\epsfig{file=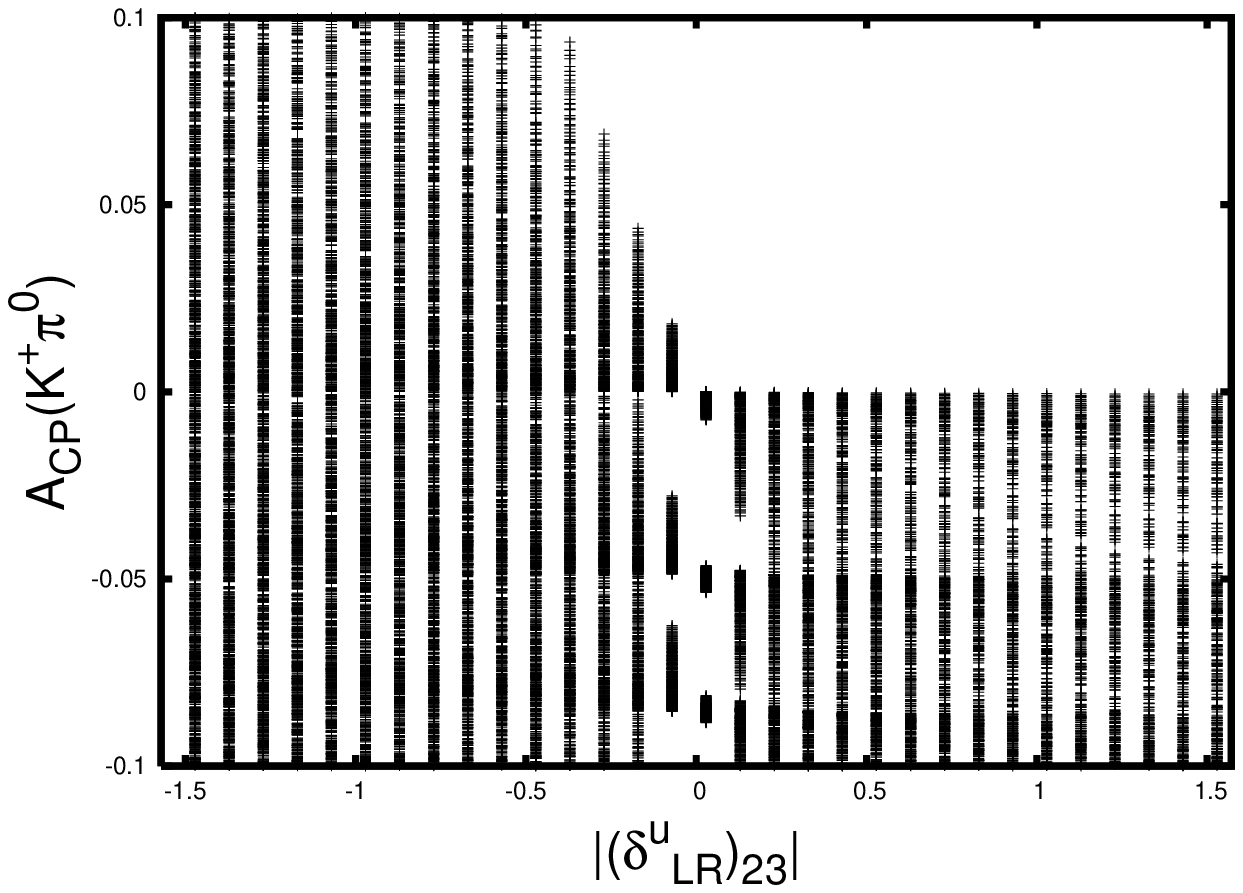,width=7cm,height=5.cm} %
\caption{(left) CP asymmetry $A_{CP}(K^+ \pi^0)$ as function of
$\vert(\delta^u_{LR})_{32}\vert$ for
$\rm{Arg}[(\delta^u_{LR})_{32}]\in[-\frac{\pi}{2},\frac{\pi}{2}]$.
(right) $A_{CP}(K^+ \pi^0)$ as function of
$\rm{Arg}[(\delta^u_{LR})_{32}]$ for
$\vert(\delta^u_{LR})_{32}\vert\in [0,1]$. The strong phases
$\delta_{\scriptscriptstyle  T}\in \left[-\pi,0\right]$ and
$\delta_{\scriptscriptstyle  EW}\in \left[0,\pi\right]$.}
\end{center}
\end{figure}

In Figs. 1 and 2, we present our numerical results for the CP
asymmetries $A_{CP}(K^+ \pi^-)$ and $A_{CP}(K^+ \pi^0)$ as
functions of the absolute value and the phase of the dominant mass
insertion $(\delta^u_{LR})_{32}$, respectively. We have scanned
over the relevant strong phases: $\delta^{\scriptscriptstyle T}
\in \left[-\pi ,0 \right]$ and $\delta^{\scriptscriptstyle EW}\in
\left[0,\pi\right]$. We have used $\gamma \simeq \pi/3$, which
gives the best fit for the SM results with the CP experimental
measurements. It turns out that $\sin(\theta_{\scriptscriptstyle
P} +\gamma)$ is usually negative, there for a negative
$\sin\delta_{\scriptscriptstyle T}$ is needed to compensate this
sign and leads to a negative $A_{CP}(K^+\pi^-)$, in agreement with
the experimental data. In our numerical analysis, the QCD
factorization approximation have been used to estimate the
hadronic matrix elements, as in Ref.\cite{Khalil:2005qg}.

From Fig. 1, one can see that within the range of input values
used for the strong phases, the predicted results of
$A_{CP}(K^+\pi^-)$ in supersymmetric models are always negative.
Thus, the experimental results at $1~\sigma$ level, \ie,
$A_{CP}(K^+\pi^-) \in \left[-0.115, -0.083\right]$ can be
naturally accommodated with $\vert (\delta^u_{LR})_{32}\vert \gsim
0.05$, and no constraint can be imposed on the phase of this mass
insertion, although negative region, \ie\ $0<
\rm{Arg}[(\delta^u_{LR})_{32}] < -\pi/2$, seems more favored.

Furthermore, Fig. 2 implies that the CP asymmetry $A_{CP}(K^+
\pi^0)$ can be in its experimental range $\left[0.025,
0.075\right]$ when $\vert (\delta^u_{LR})_{32}\vert \gsim 0.1$.
Also from the second plot in this figure, one can observe that
negative values for the phase of $(\delta^u_{LR})_{23}$ are
favored, consistently with the conclusion deduced from the result
of the asymmetry $A_{CP}(K^+\pi^-)$.

Now, let us examine the CP asymmetry $A_{CP}(K^0\pi^0)$ in this
region favored by the asymmetries $A_{CP}(K^+\pi^-)$ and
$A_{CP}(K^+\pi^0)$. It turns out that at $\vert
(\delta^u_{LR})_{23}\vert \simeq 0.4$ and
$\rm{Arg}[(\delta^u_{LR})_{23}]\simeq {\cal O}(-1)$, which lead to
consistent results for both of $A_{CP}(K^+\pi^-)$ and
$A_{CP}(K^+\pi^0)$ with their experimental measurements, one can
easily obtain $A_{CP}(K^0\pi^0) \simeq -0.1$, in agreement with
the experimental result given in Table 1.

%
\section{\large{\bf Large mixing in SUSY models }}
As shown in the previous sections, a large mixing between third
and second generation of up-squarks is required in order to
provide a solution to the $B\to K \pi$ puzzle. One may ask, is it
possible to generate such a large mixing between $LR$-squarks in
SUSY models at electroweak scale without contradicting any other
flavor changing neutral current constrains.

In fact, generally there are two ways to obtain a large $LR$
mixing that may lead to $(\delta^u_{LR})_{32}\sim {\cal O}(1)$.
The first way is through non-universal trilinear $A$-terms. In
this case,
$(\delta^u_{LR})_{23}$ is given by%
\be%
(\delta^u_{LR})_{23} \simeq \frac{1}{\tilde{m}^2} \left[
V_L^{u^+}. (Y^u A^u). V_R^u \right]_{23},%
\ee %
where $V^u_{L,R}$ are the diagonalization of the up quark mass
matrix and $\tilde{m}$ is the average squark mass. Therefore, with
a non-hierarchal Yukawa couplings and $A$-terms of order
$\tilde{m}$, it is quite plausible to obtain
$(\delta^u_{LR})_{32}$ of order ${\cal O}(1)$. However, in order
to avoid the stringent constraints from the electric dipole moment
experimental limits, the $A$-term should be Hermitian or it must
have a specific pattern. This type of non-universal $A$-terms is a
salient feature of soft SUSY breaking terms in string or brane
inspired models \cite{nonuniversal}.

The second approach for generating large $LR$  mixing is through
the non-universal soft scalar masses and large $\tan \beta$. The
simplest example of this class of SUSY models is the one suggested
in Ref.\cite{Khalil:2005dx} as a minimal of non-minimal flavor
SUSY model. In these models, the soft SUSY breaking terms are
universal except for the third generation squark masses. As
explained in Ref.\cite{Khalil:2005dx}, an effective
$(\delta^u_{LR})_{32}$ mass insertion can be obtained as follows:%
\be %
(\delta^u_{LR})_{32} \approx (\delta^u_{LL})_{32}~
(\delta^u_{LR})_{22}~, %
\ee %
where the mass insertion $(\delta^u_{LR})_{22}$ is approximately
given by $m_c A_c/\tilde{m}^2$. Here $m_c$ is the charm quark mass
and $A_c$ is the associate trilinear coupling. Therefore, in order
to have $(\delta^u_{LR})_{22} \simeq 1$, the value of $A_c$ at the
weak scale should be of order $\tilde{m}^2$, which seems
unnatural. Furthermore, since the mass insertion
$(\delta^u_{LL})_{32}$ is constrained by the branching ratio of
$b\to s \gamma$ to be $\lsim 0.1$, one finds that in this case the
resulting mass insertion  $(\delta^u_{LR})_{32}$ is less than
$0.1$. In this respect, one may conclude that the non-universal
$A$-terms is an essential requirement in order to have a
supersymmeric solution for the $B\to K \pi$ puzzle.

%
\section{\large{\bf Conclusions }}
In this letter we have updated our analysis for the supersymmetric
contributions to the branching ratios and CP asymmetries of $B\to
K \pi$ decays in the light of recent experimental measurements. We
have shown that the new experimental results for the branching
ratios allow suprsymmetry to become a natural solution for the $B
\to K\pi$ puzzle in a wider region of the parameter space. We have
found that within SUSY models with large LR mixing between second
and third generation of up-squarks, the chargino contributions can
enhance electroweak penguin and accommodate the experimental
results for both of the branching ratios and CP asymmetries of  $B
\to K\pi$. Finally we emphasized that the non-universal soft SUSY
breaking $A$-terms may be the only way to generate the required
mixing, which is necessary for this supersymmetric solution.


\end{document}